\documentclass[aps,prb,reprint,showpacs,superscriptaddress,floatfix,amsmath,amsfonts]{revtex4-1}

\usepackage{xcolor}
\definecolor{darkgreen}{rgb}{0,0.6,0}
\definecolor{cyan}{rgb}{0,0.7,0.8}
\usepackage[colorlinks=true,citecolor=darkgreen,urlcolor=cyan]{hyperref}
\usepackage{graphicx}
\usepackage{amsthm}

\theoremstyle{definition}
\newtheorem{thm}{Theorem}

\newcommand{\mrm}[1]{\mathrm{#1}}
\newcommand{\mbf}[1]{\mathbf{#1}}

\newcommand{\mc}[1]{\mathcal{#1}}

\newcommand{\eref}[1]{(\ref{#1})}
\newcommand{\Eref}[1]{Eq.~(\ref{#1})}

\newcommand{\fref}[1]{Fig.~\ref{#1}}

\newcommand{\Thref}[1]{Theorem~\ref{#1}}
\newcommand{\thref}[1]{Theorem~\ref{#1}}
\newcommand{\sref}[1]{Sec.~\ref{#1}}

\newcommand{\aref}[1]{Appendix~\ref{#1}}

\newcommand{\pfref}[1]{\protect{Fig.~\ref{#1}}}

\newcommand{\ra}{\rangle}
\newcommand{\la}{\langle}

\newcommand{\rcite}[1]{Ref.~\onlinecite{#1}}

\newcommand{\pcite}[1]{\protect{\cite{#1}}}

\newcommand{\MATLAB}{\textsc{Matlab}}
\newcommand{\ttt}[1]{\texttt{#1}}

\newcommand{\bra}[1]{\mbox{$\langle #1 |$}}
\newcommand{\ket}[1]{\mbox{$| #1 \rangle$}}

\newcommand{\T}{ \mathcal T}
\renewcommand{\L}{ \mathcal L}
\newcommand{\C}{ \mathcal C}
\renewcommand{\H}{\hat{H}}
\newcommand{\h}{\hat{h}}

\begin{document}

\title{Improving the efficiency of variational tensor network algorithms}

\author{Glen Evenbly}
\email{evenbly@caltech.edu}
\affiliation{Institute for Quantum Information and Matter, California Institute of Technology, Pasadena CA 91125, USA}
\author{Robert N. C. Pfeifer}
\email{rpfeifer@perimeterinstitute.ca}
\affiliation{Perimeter Institute for Theoretical Physics, 31 Caroline St. N, Waterloo, Ontario N2L 2Y5, Canada}  
\date{\today}

\begin{abstract}
We present several results relating to the contraction of generic tensor networks and discuss their application to the simulation of quantum many-body systems using variational approaches based upon tensor network states. Given a closed tensor network $\T$, we prove that if the environment of a single tensor from the network can be evaluated with computational cost $\kappa$, then the environment of any other tensor from $\T$ can be evaluated with identical cost $\kappa$. Moreover, we describe how the set of all single tensor environments from $\T$ can be simultaneously evaluated with fixed cost $3\kappa$. The usefulness of these results, which are applicable to a variety of tensor network methods, is demonstrated for the optimization of a Multi-scale Entanglement Renormalization Ansatz (MERA) for the ground state of a 1D quantum system, where they are shown to substantially reduce the computation time.
\end{abstract}


\maketitle

\section{Introduction}

Tensor network states have proven to be indispensable tools for the numerical simulation of quantum many body systems, and are becoming increasingly important as a framework for their theoretical understanding. 
Introduced two decades ago, White's Density Matrix Renormalization Group (DMRG) algorithm,\cite{white1992} which may be viewed as a variational algorithm for the optimization of the Matrix Product State (MPS) Ansatz,\cite{schollwock2011} remains today the preferred numerical method for one dimensional quantum systems. In more recent times new tensor network states have been introduced which offer the potential for scalable simulation of two dimensional quantum systems, including Projected Entangled Pair States (PEPS)\cite{verstraete2004,jordan2008} and the 2D Multi-scale Entanglement Renormalization Ansatz (MERA).\cite{cincio2008,evenbly2009b} However, these more recently introduced tensor network states suffer from the twin drawbacks that they are labour-intensive to implement, and computationally expensive to optimize. 
In this manuscript we present several results on the contraction of tensor networks which assist in alleviating both of these problems. 

\begin{figure}
\begin{center}
\includegraphics[width=8cm]{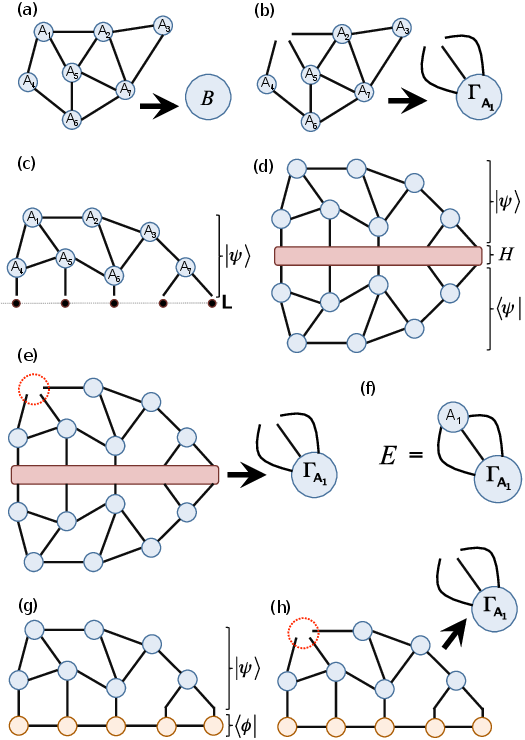}
\caption{COLOR ONLINE. (a)~An example closed tensor network, here consisting of tensors $\{A_1, A_2, A_3, A_4, A_5, A_6, A_7 \}$, evaluates to a scalar $B$ under contraction of its internal indices. (b)~An open tensor network is obtained by removing $A_1$ from the closed network in diagram~(a); this open network evaluates to a tensor $\Gamma_{A_1}$ that we call the environment of $A_1$. (c)~An open tensor network, here consisting of tensors $\{A_1, A_2, A_3, A_4, A_5, A_6, A_7 \}$, describes a quantum state $\ket{\psi}$ on the lattice $\L$. (d)~Given the Hamiltonian $\H$ defined on $\L$, the closed tensor network $\bra{\psi} \H \ket{\psi}$ evaluates to the energy $E$ of the tensor network state $\ket{\psi}$ with regards to $\H$. (e)~The environment $\Gamma_{A_1}$ of tensor $A_1$ from the closed tensor network $\bra{\psi} \H \ket{\psi}$ is evaluated. (f)~The environment $\Gamma_{A_1}$ and tensor $A_1$ are contracted to give the energy of state $\ket{\psi}$. (g)~A closed tensor network representing the overlap $\left\langle {\phi } | {\psi } \right\rangle $ of two tensor network states $\ket{\phi}$ and $\ket{\psi}$. (h)~The environment $\Gamma_{A_1}$ of tensor $A_1$ from the closed tensor network in diagram~(g) is evaluated.
\label{fig:VarPrinciple}}
\end{center}
\end{figure}

Let $\T$ be a \emph{closed} tensor network (i.e.~a network which evaluates to a scalar) on $N$ tensors $\{A_1, A_2, \ldots, A_N \}$, such as depicted in \fref{fig:VarPrinciple}(a). 
One can obtain an \emph{open} tensor network from $\T$ by removing a single tensor $A_i$ from the network, leaving the indices that were connected to this tensor unsummed. This open network evaluates to a tensor $\Gamma_{A_i}$ of the same dimension
as the removed tensor,\footnote{The dimension of a tensor is defined as the product of the dimensions of each of its indices; equivalently, this is the number of parameters defining the tensor.} 
\begin{equation}
\left| \Gamma_{A_i} \right|= \left| {A_i} \right|, 
\end{equation}
which we call the \emph{environment} of $A_i$ [\fref{fig:VarPrinciple}(b)]. The computation of such environments is crucial to the implementation of many tensor network algorithms, including variational optimization of both the PEPS and MERA Ans\"atze discussed above, and typically large numbers of environments are calculated from each closed network. In this manuscript we show that all environments computed from a single closed tensor network may be obtained at equal computational cost, and introduce a systematic and efficient method for computing multiple environment tensors from a single closed network which significantly outperforms the na\"\i{}eve calculation of each environment in turn.

We begin in \sref{sec:roleenv} by providing a more detailed description of the role played by calculation of the environment, in the context of a variational algorithm for the ground state energy of a Hamiltonian. No assumptions are made about the structure of this tensor network, which may be one-, two-, or higher-dimensional. This is followed in \sref{sec:survey} by an overview of the results which will be demonstrated in this paper, providing both context and objective for the more detailed analysis of network contraction techniques and costs presented in \sref{sec:contcosts}. Numerical demonstrations of the efficacy of these techniques are provided in \sref{sec:numeric}.

\section{Role of the environment\label{sec:roleenv}}

A common application of tensor network methods is to find (or approximate) a description of the ground state of a lattice Hamiltonian $\H\in \L$ in terms of a tensor network state $\ket{\psi}$. 
In order to approximate the ground state of $\H$, the tensors that define the state $\ket{\psi}$ will typically be chosen to minimize the energy, $E = \bra{\psi} \H \ket{\psi}$. This is commonly achieved by performing variational updates on a single tensor at a time, with the replacement tensor being chosen to minimise the energy under the assumption that the rest of the network is held constant. The single tensor update procedure is sequentially applied to all distinct tensors of the network, with this process being termed a variational sweep, and variational sweeps are iterated until the approximation to the ground state is satisfactorily converged. 

To illustrate this approach let $\L$ be a lattice of $n$ sites, each associated with a Hilbert space $\mathbb V$ of finite dimension $d$. Then an open tensor network $\T$ composed of tensors $\{A_1, A_2, \ldots, A_N \}$, with $n$ open $d$-dimensional indices, describes a quantum state $\ket{\psi}$ on $\L$; see \fref{fig:VarPrinciple}(c). 
To variationally optimize a single tensor $A_i$ one may first compute its environment $\Gamma_{A_i}$ from $\bra{\psi} \H \ket{\psi}$ [\fref{fig:VarPrinciple}(d)-(e)]. Then, recognizing that the scalar product of tensor $A_i$ with its environment $\Gamma_{A_i}$ corresponds to the energy [\fref{fig:VarPrinciple}(f)], a new tensor $A'_i$ is chosen to replace $A_i$ such that $A'_i$ minimizes the product $A'_i\,\Gamma_{A_i}$, subject to the normalization constraint $\la\psi|\psi\ra=1$.
Note that $\bra{\psi} \H \ket{\psi}$ forms a closed tensor network. 
\footnote{In optimizing the tensors of \pfref{fig:VarPrinciple}(d), one may notice that the tensors of the bra and ket in \protect{$ \bra{\psi} \H \ket{\psi}$} are related to one another via complex conjugation. However, for algorithmic convenience it is nevertheless acceptable to treat all tensors as independent degrees of freedom. As \protect{$\bra{\psi} \H \ket{\psi}$} converges towards the ground state energy of the Hamiltonian, the tensor network representations of \protect{\bra{\psi}} and \protect{$\left(\ket{\psi}\right)^\dagger$} will necessarily converge. An alternative, and slightly faster, optimization strategy is to variationally optimize only tensors in the upper (ket) portion of the diagram, and after updating a tensor in the ket portion, to replace the corresponding tensor in the bra portion with the conjugate of the newly-obtained tensor from the ket.}

Depending on the algorithm being employed the updated tensor $A'_i$ may be determined in a number of ways, for example through the singular value decomposition of $\Gamma_{A_i}$,\cite{evenbly2009} or through more sophisticated numerical techniques such as 
the method of steepest descent. In practice the precise details of the construction of $A'_i$ are unimportant to the present discussion, save for the observation that this variational optimization process
is almost entirely based upon the computation of single tensor environments from a closed tensor network (or for some algorithms, a finite set of closed tensor networks). This strategy is widely used as the preferred means of optimizing the MERA, whereas DMRG uses a slightly different, but related, strategy.\cite{schollwock2011}

Another operation often performed as part of a tensor network algorithm is the approximation of one tensor network state by another tensor network state having a different tensor structure; this type of operation may sometimes be required in imaginary-time and real-time evolution algorithms,
or in the evaluation of properties such as entanglement entropy from a tensor network state.
\cite{vidal2004,vidal2007b,verstraete2004,jordan2008,white2004,daley2004,rizzi2008,%
alba2011,koffel2012,calabrese2013,coser2013}
Given a tensor network $\T_{0}$ representing a state $\ket{\phi}$, one might seek to optimize a structurally different tensor network $\T_{1}$ to best approximate $\ket{\phi}$. That is, given that tensor network $\T_{1}$ represents a state $\ket{\psi}$, the objective is to vary the tensors in $\T_{1}$ so as to maximize the overlap $\la\phi|\psi\ra$, %
represented graphically in \fref{fig:VarPrinciple}(g). Similar to the energy minimization scheme described above, the overlap may be maximized by iteratively updating single tensors so as to increase the numerical value associated with \fref{fig:VarPrinciple}(g). Once again, a robust variational method for updating tensor $A_i$ from $\T_{1}$ may be constructed which is based around computation of the environment $\Gamma_{A_i}$ from the closed tensor network $\la\phi|\psi\ra$, as shown in \fref{fig:VarPrinciple}(h)%
.

Given the important role played by %
environment tensors in many tensor network algorithms, it is desireable that the computation of these objects $\Gamma_{A_i}$ should proceed in a computationally efficient fashion.
The results introduced in the following sections serve to simplify and make more efficient the evaluation of these single tensor environments from a closed tensor network, and thus these results have applications towards improving the performance of %
simulation algorithms for a number of different tensor network Ans\"atze. In \sref{sec:example} the usefulness %
of these results is explicitly demonstrated for optimization of a MERA representing the ground state of a 1D quantum system. %

\section{Survey of results\label{sec:survey}} 

Let $\T$ be a closed tensor network of $N$ tensors $\{A_1, A_2, \ldots, A_N \}$. The main results of this manuscript are as follows:
\begin{thm}
	We prove that if the environment $\Gamma_{A_i}$ of one tensor $A_i$ can be evaluated from $\T$ with a total computational cost $\kappa$, then the environment $\Gamma_{A_j}$ of any other tensor $A_j$ from $\T$ can be computed with \emph{exactly} the same cost $\kappa$. This proof is constructive: if the contraction order for evaluating the environment $\Gamma_{A_i}$ with computational cost $\kappa$ is known, then we describe how contraction orders can be identified for the evaluation of any other environment $\Gamma_{A_j}$ for $1\le j\le N$ with the same cost $\kappa$. It follows that if $\kappa_\mrm{min}$ is the minimal cost for evaluating $\Gamma_{A_i}$, i.e. the cost resulting from contraction according to some optimal sequence, then this is also the minimal cost for evaluating any other environment $\Gamma_{A_j}$ for $1\le j\le N$. (The problem of identifying the optimal cost $\kappa_\mrm{min}$ and an associated contraction sequence is addressed in Ref.~\onlinecite{pfeifer2013a}.) \label{result1}
\end{thm}
\begin{thm}
	If the environment $\Gamma_{A_i}$ of one tensor $A_i$ can be evaluated  with a total computational cost $\kappa_\mrm{min}$, we show that the set of all single tensor environments $\{\Gamma_{A_1}, \Gamma_{A_2}, \ldots, \Gamma_{A_N} \}$ can be simultaneously evaluated from $\T$ with a total computational cost of $3\kappa_\mrm{min}$. Again, this result is constructive. By comparison, if one were to compute each environment separately then \thref{result1} would imply a na\"\i{}eve total cost of $N\kappa_\mrm{min}$. \label{result2}
\end{thm}

\section{Contraction of tensor networks\label{sec:contcosts}} 

\subsection{Overview}

\begin{figure}
\begin{center}
\includegraphics[width=8cm]{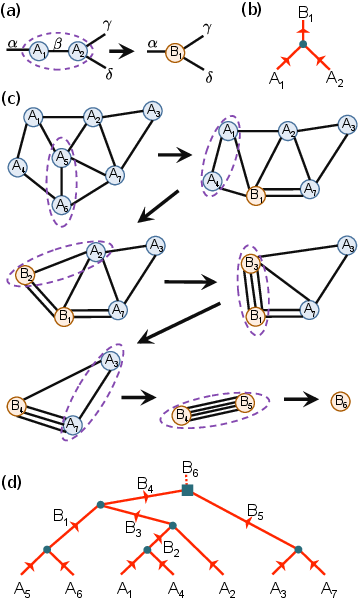}
\caption{COLOR ONLINE. (a)~Two tensors $A_1$ and $A_2$ are contracted to give a new single tensor $B_1$. The indices labelled $\alpha$, $\beta$, $\gamma$, and $\delta$ have dimensions $|\alpha|$, $|\beta|$, $|\gamma|$, and $|\delta|$ respectively.
Following \Eref{eq:1}, the computational cost of performing this contraction is $\mbf{cost}\!:\!(A_1,A_2)=\chi_{13} \chi_{12} \chi_{24} \chi_{25}$. (b)~A contraction tree representing the tensor contraction from diagram~(a). (c)~A closed tensor network $\T$ is contracted to a scalar $B_6$ following a sequence of pairwise tensor contractions (indicated by dashed ovals). (d)~Representation of the tensor contractions of diagram~(c) as a contraction tree. Edges of the graph represent tensors, and vertices of the tree represent the contraction of two tensors to give a new tensor. The open edges represent the original tensors of the network $\T$, while internal edges represent the tensors obtained during intermediate stages of the contraction. The root vertex, depicted as a square, represents the contraction of two tensors into a scalar, denoted $B_6$.
\label{fig:Contract}}
\end{center}
\end{figure}

Preparatory to proving the results of the previous section, it is first necessary to discuss the practical means by which a closed tensor network may be contracted to a scalar and the computational cost of doing so. It can be shown that the optimal strategy for contracting a tensor network is through a sequence of pairwise contractions,\cite{pfeifer2013a} where each pairwise contraction involves summing the indices connecting two tensors to obtain a new tensor. We use a bracket notation to denote the pairwise contraction of two tensors, such that $B_1=(A_1,A_2)$ represents the contraction of tensors $A_1$ and $A_2$ to give a new tensor $B_1$ [\fref{fig:Contract}(a)]. 
It is frequently convenient to express a pairwise tensor contraction in the form of a matrix multiplication, combining all indices not being summed over into a single index on each tensor, and similarly for all indices being summed over. In the example of \fref{fig:Contract}(a) tensor $A_1$ has two indices and tensor $A_2$ has three indices, labelled as shown in the figure, and contraction of these two tensors corresponds to evaluation of the product
\begin{equation}
\left({B}_1\right)_{\alpha\gamma\delta}=\sum_{\beta} \left({A}_1\right)_{\alpha\beta}\left({A}_2\right)_{\beta\gamma\delta}.
\end{equation}
Combining indices $\gamma$ and $\delta$ (of dimensions $|\gamma|$ and $|\delta|$ respectively) into a single index $\epsilon$ of dimension $|\gamma|\times|\delta|$ permits this process to be written in the form of a matrix multiplication
\begin{equation}
\left({B}_1\right)_{\alpha\epsilon}=\sum_{\beta} \left({A}_1\right)_{\alpha\beta}\left({A}_2\right)_{\beta\epsilon}.
\end{equation}
The computational cost of performing such a contraction typically scales as the number of multiplication operations which must be performed, here $|\alpha|\times|\beta|\times|\epsilon|$. More generally, for two arbitrary tensors $A_i$ and $A_j$, if the total dimension of all indices connecting these two tensors is denoted $\chi_{ij}$, then it follows that
\begin{equation}
{\mbf{cost}}\!:\!\left( {{A_i},{A_j}} \right) = \frac{\left|A_i \right| \left|A_j \right|}   {\chi _{ij}}. \label{eq:1}
\end{equation}

Any closed tensor network of $N$ tensors can be reduced to a scalar through a series of $N-1$ pairwise contractions; in \fref{fig:Contract}(c), for example, a closed tensor network of seven tensors $A_1,\ldots,A_7$ is reduced to a scalar $B_6$ through the sequence of pairwise contractions
\begin{equation}
{B_6} = \left( {\left( {\left( {{A_5},{A_6}} \right),\left( {\left( {{A_1},{A_4}} \right),{A_2}} \right)} \right),\left( {{A_3},{A_7}} \right)} \right).
\end{equation}
As is discussed in \rcite{pfeifer2013a}, the total cost for contracting a tensor network such as this will in general depend on the sequence in which these pairwise contractions are performed,
with some sequences being computationally more expensive than others. Finding an optimal sequence of contractions, i.e.~one with the smallest possible computational cost, is a difficult problem which is frequently best handled by automated search algorithms.\cite{pfeifer2013a}

When considering the evaluation of a given tensor network, it is useful to represent a particular sequence of pairwise contractions in the form of a \emph{contraction tree} as shown in \fref{fig:Contract}(b)-(d). A contraction tree is a rooted, unbalanced binary tree (a binary tree where the depth of the subtrees from each node may differ), with arrows directing each edge from the child to the parent node (i.e.~describing a flow \emph{towards} the root, the exact opposite of an arborescence).
Edges of the tree are representative of tensors, with edges at maximal depth representing the initial tensors of the network and internal edges representing tensors obtained in intermediate stages of the contraction. Each vertex of the tree represents the contraction of a pair of tensors; the root vertex of the tree represents the contraction of two tensors into the scalar, while all other vertices, which we refer to as ternary vertices, represent the contraction of the two tensors represented by the incoming edges into the new tensor represented by the outgoing edge.   

\subsection{Three tensor permutation relation}
\begin{figure}
\begin{center}
\includegraphics[width=8cm]{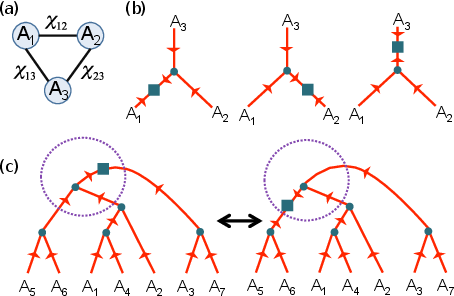}
\caption{COLOR ONLINE. (a)~A closed tensor network $\T$ of three tensors $\{ A_i, A_2, A_3\}$, %
with $\chi_{ij}$ representing the dimension of the index connecting tensors $A_i$ and $A_j$. (b)~Contraction trees representing three different ways to contract the network of diagram~(a) to a scalar. Each of these contraction orders has the same leading order cost [see \Eref{eq:TotCost}]---the costs differ only in the contraction at the root vertex, which is never greater than that associated with the preceding ternary vertex. (c) Two contraction trees that differ only within the dotted circles (relating to the contraction order of their final three tensors). The result from diagram~(b) implies that the cost associated with each of these contraction orders differs only in the contraction at the root vertex (represented by the square).
\label{fig:Triangle}}
\end{center}
\end{figure}

Consider a closed tensor network $\T$ of three tensors $\{ A_1, A_2, A_3 \}$, as shown in \fref{fig:Triangle}(a), where tensor pairs $A_i$ and $A_j$ are connected by a single index of dimension $\chi_{ij}$. 
This network can be contracted to a scalar (through a sequence of two pairwise contractions) in three non-equivalent ways. The total computational costs of these contractions are:
\begin{equation}
\begin{split}
  {\mbf{cost}}\!:\!\left( {\left( {{A_1},{A_2}} \right),{A_3}} \right) &= {\chi _{12}}{\chi _{13}}{\chi _{23}} + {\chi _{13}}{\chi _{23}} \hfill \\
  {\mbf{cost}}\!:\!\left( {\left( {{A_1},{A_3}} \right),{A_2}} \right) &= {\chi _{12}}{\chi _{13}}{\chi _{23}} + {\chi _{12}}{\chi _{23}} \hfill \\
  {\mbf{cost}}\!:\!\left( {\left( {{A_2},{A_3}} \right),{A_1}} \right) &= {\chi _{12}}{\chi _{13}}{\chi _{23}} + {\chi _{12}}{\chi _{13}} \hfill.  \label{eq:TotCost}
\end{split}
\end{equation}
Notice that the three contractions orders, whose corresponding contraction trees are depicted in \fref{fig:Triangle}(b), only differ in the cost of the second contraction which brings two tensors into a scalar, and that these costs are always less than or equal to the cost of the first contraction. The cost of the first pairwise contraction is the same across all three contraction orders:
\begin{equation}
\begin{split}
{\mbf{cost}}\!:\!\left( {{A_1},{A_2}} \right) = {\mbf{cost}}\!:\!\left( {{A_1},{A_3}} \right) &={\mbf{cost}}\!:\!\left( {{A_2},{A_3}} \right) \\&= {\chi _{12}}{\chi _{13}}{\chi _{23}}. \label{eq:SameCost}
\end{split}
\end{equation}
We call this result, which allows us to change the contraction order of closed network of three tensors without affecting the leading order contraction cost, the \emph{three tensor permutation relation}; this relation will be of key importance in deriving the main results of this manuscript.

\subsection{Contraction tree families}

In this section we use the three tensor permutation relation to derive a similar result for networks with a larger number of tensors $N$. This result is then used to establish a notion of a contraction tree \emph{family}. 

Let $\C$ be a contraction tree that contracts a network $\T$ of $N$ tensors to a scalar for some cost $\kappa+\epsilon$ (where $\epsilon$ is the cost of the final contraction of tree $\C$, associated with the root vertex). We choose one of the vertices adjacent to the root vertex to represent the penultimate contraction performed during the contraction sequence described by tree $\C$. Now, let us change the order of the final two contractions in $\C$ (which together contract three tensors to a scalar) in order to obtain a different contraction tree $\C'$. An example is given in \fref{fig:Triangle}(c); notice that this change in contraction order can be envisioned as `shifting' the location of the root vertex on the contraction tree from one edge of a trivalent vertex to another. 

By the three tensor permutation relation \eref{eq:TotCost}, the cost of contracting $\T$ in accordance with $\C'$ is $\kappa+\epsilon'$ (where $\epsilon'$ is the cost of the final contraction of tree $\C'$). Comparing this with the cost of $\kappa+\epsilon$ for tree $\C$, the cost of contracting $\T$ according to either $\C$ or $\C'$ is thus seen to differ only by the costs of the final contractions associated with the root vertices. (Notice also that this cost is necessarily less than or equal to that of the preceding contraction represented by the trivalent vertex of the contraction tree.)

This shifting of the root vertex may be applied repeatedly in order to obtain further different contraction trees. We say that any two contraction trees %
which are
related by %
one or more shifts of the root vertex, such as $\C$ and $\C'$ described above, belong to the same \emph{family}. Since a contraction tree for a tensor network of $N$ tensors has $2N-3$ distinct edges on which the root vertex may be positioned, there are $2N-3$ contraction trees $\{ \C_1, \C_2, \ldots, \C_{2N-3}\}$ in each such family. By the arguments given above, each tree $\C_i$ %
describes a contraction of network $\T$ to a scalar for total cost $\kappa+\epsilon_i$, where $\kappa$ is a fixed cost specific to the family and $\epsilon_i$ is the cost for the final (root vertex) contraction of tree $\C_i$. %
The value of $\epsilon_i$ may be different for each member of the family.

\subsection{Proof of \protect{\Thref{result1}}}

\begin{figure}
\begin{center}
\includegraphics[width=8cm]{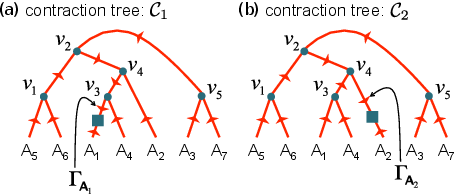}
\caption{COLOR ONLINE. (a) A contraction tree $\C_1$ that could be used to evaluate the environment $\Gamma_{A_1}$ of tensor $A_1$ (by performing all contractions excluding that associated with the root vertex). (b) A contraction tree $\C_2$ that could be used to evaluate the environment $\Gamma_{A_2}$ of tensor $A_2$. Since both contraction trees are in the same family, it follows that the cost of contracting for $\Gamma_{A_1}$ is identical to that of $\Gamma_{A_2}$. Notice that the contractions associated with vertices $\mc{V}_1$, $\mc{V}_2$ and $\mc{V}_5$ represent identical binary tensor contractions in both $\C_1$ and $\C_2$; it follows that, in the evaluation of both $\Gamma_{A_1}$ and $\Gamma_{A_2}$ together, these binary tensor contractions need only be performed once and the result may be reused.
\label{fig:TwoTree}}
\end{center}
\end{figure}

With the notion of \emph{families} of contraction trees now established, the proof of \thref{result1} %
follows easily. Consider a closed tensor network $\T$ of $N$ tensors $\{ A_1, A_2, \ldots, A_N \}$ and assume $\C_1$ is a contraction tree which performs a contraction of the network to yield environment $\Gamma_{A_1}$ at a cost $\kappa$, before finally contracting tensor $A_1$ with its environment, $\left( A_1, \Gamma_{A_1} \right)$. This implies that the root vertex of $\C_1$ is directly connected to the edge associated with tensor $A_1$, as in the example depicted in Fig.\ref{fig:TwoTree}(a). The final contraction at the root vertex, which evaluates the scalar of the closed network, does not need to be performed if one is interested only in obtaining $\Gamma_{A_1}$. Now let $\C_i$ be the contraction tree in the same family as $\C_1$ where the root vertex is directly connected to the edge associated with tensor $A_i$. In the contraction order specified by $\C_i$ the final step in contracting the network to a scalar would be $\left( A_i, \Gamma_{A_i} \right)$, and thus the cost of obtaining $\Gamma_{A_i}$, which arises from performing all of the contractions except that at the root vertex, is also exactly $\kappa$ (as, by assumption, the contraction trees $\C_1$ and $\C_i$ belong to the same family). Since this argument holds for any tensor $A_i$, it proves \thref{result1}%
: If the environment of one tensor can be constructed for a computational cost $\kappa$ then any of the other environments may also be constructed for the same cost $\kappa$. 

Moreover, if one identifies an optimal (lowest cost) contraction tree $\C^\mrm{opt}_i$ for evaluating one particular single tensor environment, $\Gamma_{A_i}$, with the cost of this evaluation being $\kappa_\mrm{min}$, then this result guarantees that there exists a contraction tree in the same family as $\C^\mrm{opt}_i$ which gives an optimal contraction order for any other environment $\Gamma_{A_j}$, also at cost $\kappa_\mrm{min}$. This proof is in two parts: First, by the argument presented above there necessarily exists a tree $\C^\mrm{opt}_{j}$ in the same family as $\C^\mrm{opt}_i$ which permits the evaluation of $\Gamma_{A_j}$ for cost $\kappa_\mrm{min}$. Second, suppose there existed a contraction tree $\C_j^{\prime\,\mrm{opt}}$ permitting calculation of $\Gamma_{A_j}$ for a cost $\kappa'_\mrm{min}<\kappa_\mrm{min}$. Then there would necessarily exist a tree in the same family as $\C^{\prime\,\mrm{opt}}_j$ permitting calculation of $\Gamma_{A_i}$ for cost $\kappa'_\mrm{min}$, in contradiction with the earlier statement that $\C_i^\mrm{opt}$ represents an optimal (minimum cost) contraction sequence for the construction of $\Gamma_{A_i}$. Consequently $\kappa_\mrm{min}$ is necessarily also the minimum cost for construction of $\Gamma_{A_j}$, and thus $\C^\mrm{opt}_j$ is necessarily an optimal tree for construction of $\Gamma_{A_j}$.

\subsection{Proof of \protect{\Thref{result2}}\label{sec:effsim}}

During a single iteration of a variational optimization algorithm, in which all unique tensors of the tensor network state are updated, one typically computes multiple environments from the same closed tensor network (or set of closed tensor networks). In this section we show how, by recycling the intermediate tensors that arise in the contractions, multiple environments can be efficiently computed in accordance with \thref{result2}. %

Let $\T$ be a closed tensor network composed of $N$ tensors $\{A_1, A_2, \ldots , A_N \}$, and let $\C_1$ be a contraction tree which %
yields environment $\Gamma_{A_1}$ at a cost $\kappa$. As in the previous section, we define $\C_i$ to be the contraction tree in the same family as $\C_1$ where the root vertex is directly connected to the edge associated with tensor $A_i$. Imagine that we would like to compute the set of environments for all tensors $\{\Gamma_{A_1}, \Gamma_{A_2}, \ldots , \Gamma_{A_N} \}$; na\"\i{}evely this could be achieved with total cost $N\kappa$ by contracting $\T$ using each of the contraction trees $\{ \C_1, \C_2, \ldots, \C_N \}$. In practice, however, identical binary tensor contractions will occur repeatedly during the evaluation of the different contraction trees $\C_i$, as shown for example in \fref{fig:TwoTree}, and the total cost of the set of contractions may be significantly reduced if repetition of previously-performed tensor contractions is avoided. One strategy to achieve this reduction is as follows: Let us imagine contracting $\T$ according to each tree in the family $\{ \C_1, \C_2, \ldots, \C_N \}$ in sequence. The intermediate tensors resulting from each contraction are stored in memory, and each binary tensor contraction is performed only if it has not already been performed in one of the preceding contractions of $\T$ (otherwise the result of the contraction is reused from memory). We now argue that, by following this strategy of recycling intermediate tensors, the complete set of environments $\{\Gamma_{A_1}, \Gamma_{A_2}, \ldots , \Gamma_{A_N} \}$ may be evaluated for a total cost of $3\kappa$.  

In order to compute the cost of evaluating all environments (while employing intermediate tensor recycling) we must first identify all unique contractions which appear when evaluating tensor network $\T$ according to the set of trees $\{ \C_1, \C_2, \ldots, \C_N \}$. Let the $N-2$ trivalent vertices in tree $\C_1$ be labeled $\mc{V}_1,\ldots,\mc{V}_{N-2}$. As each contraction tree in the family may be thought to differ only in the location of the root vertex and in the orientations of the directed edges, we may therefore also use the same labels $\mc{V}_1,\ldots,\mc{V}_{N-2}$ to label corresponding trivalent vertices in every member of $\{ \C_1, \C_2, \ldots, \C_N \}$; again see Fig.\ref{fig:TwoTree}. In each tree $\C_i$ a given vertex $\mc{V}_k$ will have two incoming edges and one outgoing edge, where the configuration of these orientations may vary between trees. If a trivalent vertex $\mc{V}_k$ has the same configuration in two of these trees $\C_i$ and $\C_j$ then it is seen to represent the same binary tensor contraction in each instance. This follows from the identification of contraction trees as directed acyclic graphs: if the edges meeting at a trivalent vertex $\mc{V}_k$ have the same orientations in two trees $\C_i$ and $\C_j$, then the portions of the trees connected to the inbound edges must likewise exhibit the same orientations. Thus these branches will correspond to the same sequences of tensor contractions, yielding the same tensors on the inbound edges of $\mc{V}_k$. It therefore follows that if the orientations of the edges meeting at $\mc{V}_k$ match, then vertex $\mc{V}_k$ represents exactly the same pairwise tensor contraction in both $\C_i$ and $\C_j$.

Each vertex $\mc{V}_1,\ldots,\mc{V}_{N-2}$ appears in each of its three possible configurations in at least one of the trees $\{ \C_1, \C_2, \ldots, \C_N \}$, and thus the set of unique binary contractions (required to evaluate all single tensor environments) are those corresponding to the trivalent vertices $\mc{V}_1,\ldots,\mc{V}_{N-2}$ in each of their three configurations of incoming and outgoing edges. 

Now that we have identified the unique contractions which appear when evaluating $\T$ according to the set of trees $\{ \C_1, \C_2, \ldots, \C_N \}$, it remains only to compute the cost of evaluating these unique contractions. Whereas calculation of a single environment involves the contraction of each trivalent vertex in a single configuration for a cost of $\kappa$, the calculation of all environments $\{\Gamma_{A_1}, \Gamma_{A_2}, \ldots , \Gamma_{A_N} \}$, requires evaluation of each trivalent vertex in each of the three possible configurations.
By the three tensor permutation relation \eref{eq:TotCost} the cost of evaluating the binary contraction associated with a given trivalent vertex $\mc{V}_k$ is independent of its configuration of incoming and outgoing edges, and the cost of computing all of the environments is therefore exactly $3\kappa$.
Note that if one was interested in only computing some, and not all, of the single tensor environments from network $\T$ then this could be achieved for a cost between $\kappa$ and $3\kappa$.

\section{Example application: 1D MERA\label{sec:example}}

Let us now examine how \thref{result2} %
may be applied to improve the efficiency and performance of variational tensor network algorithms, both in general and for the specific case of the binary 1D MERA. 

Assume $\ket{\psi}$ is a tensor network state on a lattice $\L$, consisting of $N$ tensors $\{ A_1, A_2, \ldots ,A_N\}$ that we want to optimize to approximate the ground state of a Hamiltonian $\H$ on $\L$. Let us consider two different optimization strategies:
\begin{enumerate}
\item[(i)] In the first strategy, computation of an environment $\Gamma_{A_i}$ is followed immediately by update of tensor $A_i$ using this environment. The calculation of subsequent environments then makes use of the updated version of tensor $A_i$. A variational sweep consists of the consecutive iteration of this procedure for $i$ ranging from 1 to $N$.
\item[(ii)] In the second strategy, one computes the entire set of environments $\{ \Gamma_{A_1},\Gamma_{A_2},\ldots, \Gamma_{A_N} \}$ simultaneously, then updates all tensors $\{A_1, A_2, \ldots, A_N \}$ simultaneously.
\end{enumerate}
Using the first strategy the cost per variational sweep may be as high as $N\kappa$, with $\kappa$ being the cost for computing a single environment. Using the second strategy, \thref{result2} implies a cost per sweep of at most $3\kappa$, potentially reducing the computation time per sweep by $O(N)$. 

Although this cost saving is significant, the simultaneous evaluation of environments employed in approach~(ii)
is not without potential drawbacks. 
During the sequential update procedure described in approach~(i), the environment computed for a given tensor $A_i$ incorporates changes already made to tensors other than $A_i$ earlier in the sweep, whereas it is not in general feasible to incorporate such changes into environments under the simultaneous update approach. One may therefore expect a tensor network algorithm to require fewer iterations to converge when employing the sequential update approach, and also to converge more smoothly as fewer parameters are changed at once. In addition, more memory is required for storage of intermediate tensors during simultaneous computation of environments and this could be significant in memory-limited calculations.

We compare the two strategies for the optimization of a 1D binary MERA for a 1D Hamiltonian 
\begin{equation}
\H=\sum_r \h^{[r,r+1,r+2]} %
\end{equation}
where $\h^{[r,r+1,r+2]}$ is a 3-site operator acting on lattice sites $r$, $r+1$, and $r+2$, and $\H$ is
chosen to be translation-invariant (i.e.~all $\h^{[r,r+1,r+2]}$ identical). A translation-invariant MERA is used to approximate the ground state, in which each layer is defined by two distinct tensors (an isometry $w$ and disentangler $u$). The standard optimization sweep for MERA,\cite{evenbly2009} which proceeds layer by layer, is already structured in such a way as to recycle intermediate tensors from different layers of the sweep. (Specifically, this is achieved by computing the effective Hamiltonian couplings and local reduced density matrices at a given level $\tau$ of coarse graining, denoted $h_\tau$ and $\rho_\tau$ respectively in \rcite{evenbly2009}.)

\begin{figure}
\begin{center}
\includegraphics[width=8cm]{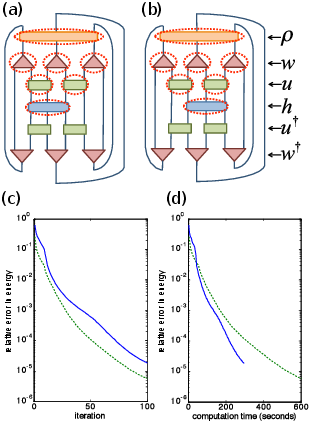}
\caption{(a)-(b)~The two closed tensor networks that arise in the optimization of a 1D binary MERA.\pcite{evenbly2009} The dotted ovals indicate the single tensor environments that need to be computed during each iteration of the optimization. (c)~Plot of error in the ground state energy against iteration number during optimization of a binary MERA for the ground state of the critical Ising model on a 1D lattice of 72 sites. The solid line represents the new algorithm where all environments are computed simultaneously, whereas the dashed line represents the traditional algorithm where environments are computed sequentially. The algorithm based upon simultaneous computation of environments takes more iterations to converge to the same energy. (d)~In this diagram the errors in ground state energy from diagram~(c) are plotted against time, and the advantages of the new algorithm become apparent. Although the original algorithm requires less iterations to converge, these iterations take significantly longer than those of the new algorithm, with the new algorithm requiring approximately $40\%$ less computing time than the original algorithm to reach the chosen error threshold of one part in $10^{-5}$. \label{fig:MERA}}
\end{center}
\end{figure}

Given that the standard MERA algorithm already recycles intermediate tensors (to a limited degree) between layers during the sweep, one can not expect to achieve a performance increase of $O(N)$ on applying the simultaneous update strategy. Nevertheless, \thref{result2} %
can still be used to reduce the computation time for environments within each layer. The optimization of tensors $w$ and $u$ in any given layer of the MERA involves the calculation of environments which arise from two distinct closed tensor networks, depicted in \fref{fig:MERA}(a)-(b). Environments are constructed for each non-conjugated instance of the disentangler $u$ (two environments per diagram) and the isometry $w$ (three environments per diagram), while the environments of the three-site Hamiltonian coupling and reduced density matrix also need to be computed for later use. (The calculation of these environments corresponds to the ``lowering'' of $\rho_{\tau+1}$ to lattice $\L_\tau$ and the ``lifting'' of $h_\tau$ to lattice $\L_{\tau+1}$ respectively, as described in \rcite{evenbly2009}.) Overall this requires the calculation of seven different single tensor environments from each of the closed tensor networks, as indicated in \fref{fig:MERA}(a)-(b).

Figures~\ref{fig:MERA}(c)-(d) compare convergence rate and computation time for the 1D critical Ising model under the two different optimization strategies. In \fref{fig:MERA}(c) it is seen that, in order for the Ansatz to converge to a given level of accuracy, the implementation performing simultaneous computation of environments requires more iterations than does the implementation performing sequential computation. %
However, since each iteration of the simultaneous computation algorithm takes roughly half the time of each iteration of the sequential algorithm, the overall computation time required for the simultaneous algorithm to reach the chosen accuracy threshold is only about 60\% of that required by the sequential algorithm [as shown in \fref{fig:MERA}(d)]. This performance benefit was sustained over repeated trials.

\section{Discussion and conclusions\label{sec:numeric}} 

The ability to efficiently contract tensor network Ans\"atze is fundamental to the efficacy of many variational tensor network algorithms. The results derived in this manuscript, which are applicable to arbitrary tensor networks, have the potential to simplify the implementation of tensor network algorithms and to improve their efficiency. %

\Thref{result1} %
simplifies the implementation of tensor network algorithms by substantially reducing the number of optimal contraction sequences which must be found by manual or automated search. Identifying the optimal contraction order for a given tensor network is a difficult task; typically one must resort to a brute force approach. With \thref{result1}, once the optimal contraction order for one environment from a closed network is known then the optimal contraction order for any other environment deriving from the same closed network may be obtained %
directly. This result also provides a way to check for inefficiencies in existing tensor network codes: if two environments from the same closed network are not being contracted for the same cost then this implies an inefficiency which may be corrected.

\Thref{result2} %
allows multiple environments to be efficiently computed from a closed tensor network by systematically recycling the tensors which are calculated as intermediate steps (as demonstrated, for example, in \sref{sec:example} and implemented in \aref{sec:RefImp}). 
This approach is capable of providing significant increases in computational performance for variational tensor network algorithms: Given a tensor network state consisting of $N$ distinct tensors, the resulting reduction in time required for a variational sweep may be a factor as large as $O(N)$. In practice, however, many existing tensor network algorithms already recycle some intermediate tensors, and so the full $O(N)$ performance increase is seldom attained%
: In %
DMRG %
the left and right blocks are recycled, and likewise in MERA %
the effective Hamiltonians and reduced density matrices at different levels of coarse-graining are recycled. %
While the approach of \thref{result2} does not in these instances confer a speedup that scales with $N$, the total number of tensors in the Ansatz, it is nevertheless still useful in maximizing the extent to which recycling of intermediate tensors takes place. This was
demonstrated for the binary 1D MERA in \sref{sec:example} where an overall performance increase of $40\%$ (independent of $N$) was observed. Moreover, the approach presented in this paper may be thought of as providing a systematic way of taking the recycling intermediate tensors, as currently exploited to some degree in the DMRG and MERA algorithms, and extending this to the variational optimization of arbitrary tensor networks.

Finally, we consider the consequences of these results for the tensor network and condensed matter communities. First, there is the obvious benefit that these results reduce the level of complexity involved in programming a tensor network algorithm. As an example, the original algorithm\cite{evenbly2009} for the binary MERA presented in \sref{sec:example} requires the programming of fourteen different tensor network contractions, whereas exploitation of the results of \sref{sec:effsim} reduces the number of networks to only two. This benefit is even more pronounced for more complex, two-dimensional Ans\"atze such as the 2D MERA or PEPS. Second, there is the ability to increase existing algorithm performance, albeit at the expense of greater memory requirements. In practice the majority of tensor network algorithms are seldom memory-limited, and thus this increase in performance corresponds to an increase in the simulation complexity which may be tackled, and thus the precision of the results which are obtained. Finally, and perhaps most profoundly, there is the extension of intermediate result recycling to arbitrarily-structured tensor networks. Existing variational tensor network algorithms generally exhibit highly ordered structures specifically designed to facilitate this recycling (for example the periodic structure of the MPS and of each layer of the MERA, and the construction of the latter from predominantly unitary and isometric tensors). By extending this recycling behaviour to arbitrarily-structured tensor networks, we open the door to efficient calculations on tensor networks of arbitrary (and perhaps even evolving) design. Coupled with the automated determination of the optimal contraction sequence for a given tensor network,\cite{pfeifer2013a} this result opens the door to an entirely new form of variational Ansatz. Tensor network Ans\"atze of this form find a natural application in the field of loop quantum gravity, where a spin network may be thought of precisely as an SU(2)-symmetric tensor network whose connections evolve over time. It is also interesting to speculate that the condensed matter tensor network algorithms of the future might display adaptive capabilities, being capable of varying their interconnections in order to better represent the entanglement structure of the state being studied.

\begin{acknowledgments}
G.E. is supported by the Sherman Fairchild foundation. R.N.C.P. thanks the Ontario Ministry of Research and Innovation Early Researcher Awards for financial support. Research at Perimeter Institute is supported by the Government of Canada through Industry Canada and by the Province of Ontario through the Ministry of Research and Innovation.
\end{acknowledgments}

\appendix

\section{Reference implementation\label{sec:RefImp}}

It is relatively straightforward to implement the approach described in \sref{sec:effsim} for an arbitrary tensor network, permitting calculation of an arbitrary number of tensor environments for a cost less than or equal to $3\kappa$ through the recycling of all relevant intermediate results. 
A reference implementation in \MATLAB{} has included with the arXiv version of this paper, and may be downloaded as described in \sref{sec:obtain}. Standard usage is described in \sref{sec:usage}.

\subsection{Obtaining the reference implementation\label{sec:obtain}}

\begin{enumerate}
\item While viewing the abstract page for the latest version of this paper (arXiv:1310.8023), click ``Download: Other formats''.
\item Click ``Download source''.
\item Save the resulting file with extension ``.tar.gz''. This file is an archive containing both the reference implementation and the \LaTeX{} source for the present paper.
\item Unpack the archive using your preferred unarchiver (on a UNIX system you could use \texttt{tar xvfz filename.tar.gz}).
\end{enumerate}
The reference implementation for efficiently computing multiple tensor environments is provided by the file named \texttt{multienv.m}. %

\subsection{Using the reference implementation\label{sec:usage}}

The reference implementation of the approach to computing multiple tensor environments described in \sref{sec:effsim} is provided in the form of a \MATLAB{} function \ttt{multienv()}. Invocation takes the form
\begin{align*}
\ttt{[env1 env2 \ldots]~=} &~\ttt{multienv(tensorList,envList,}\\&\ttt{~~~~~~~~~~legLinks,sequence);}
\end{align*}
where the input parameters are specified as follows:

\ttt{tensorList} is a $1\times n$ cell array containing the $n$ tensors which make up the tensor network.

\ttt{envList} is a $1\times n$ array of integers. If an entry in \ttt{envList} is non-zero then the environment of the corresponding tensor is computed and returned in the specified output variable. For example, if the tensor list is \ttt{\{A,B,C,D,E\}} and \ttt{envList} is \ttt{[0 2 0 0 1]} then \ttt{multienv()} will return two tensors \ttt{env1} and \ttt{env2} corresponding to the environments of tensors \ttt{E} and \ttt{B} respectively. If an integer is repeated, the corresponding environments are added together. Thus an \ttt{envList} of \ttt{[0 1 0 0 1]} would cause \ttt{multienv()} to return a single tensor, being the sum of the environments of \ttt{B} and \ttt{E}.

\begin{figure}
\includegraphics[width=246.0pt]{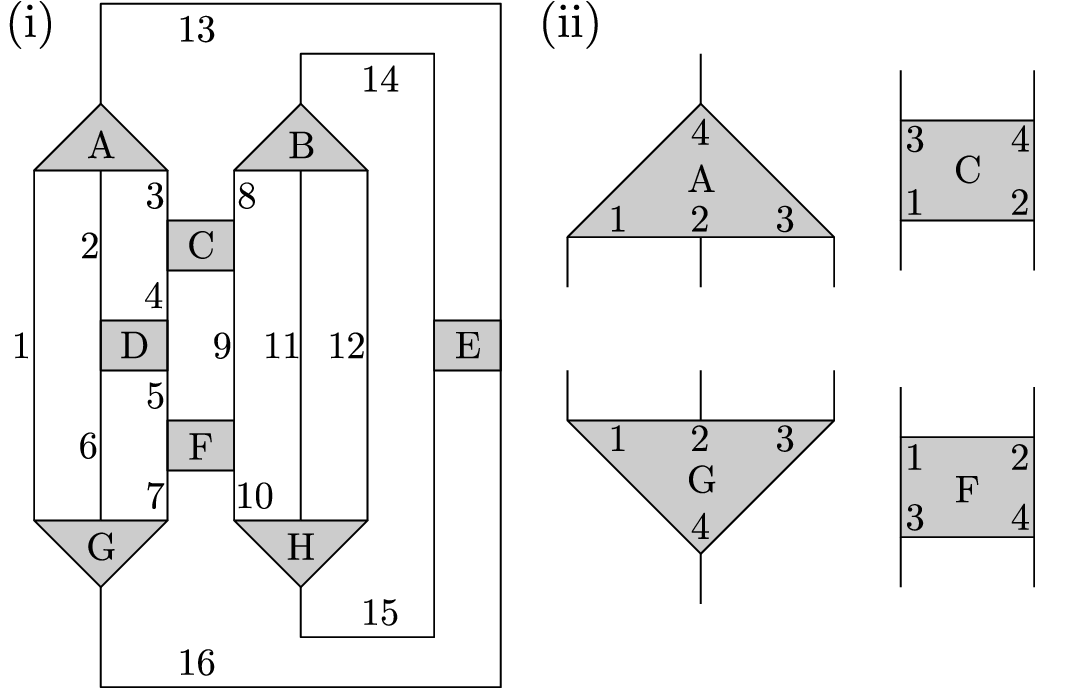}
\caption{(i)~A closed tensor network diagram arising in the optimization of the 3:1 1D MERA. (ii)~Ordering of indices on the tensors of diagram~(i). Ordering for B is the same as for A. Ordering for D and E is the same as for C. Ordering for H is the same as for G. Note that ordering for tensor F differs from C, D, and E because tensor F is customarily obtained from tensor C in the 3:1 1D MERA by complex conjugation and vertical reflection, and the process of reflection affects the leg ordering. %
\label{fig:labelledMERA}}
\end{figure}

\ttt{legLinks} describes the tensor network using the leg-labelling notation given in \rcite{pfeifer2013a}. In brief, the tensor network is represented using the customary diagrammatic notation (for which a summary may be found in \rcite{pfeifer2011b}) and an integer label is assigned to each index (represented in the diagram by a leg). Summed indices are associated with positive integer labels, while open indices are associated with negative integer labels. In the present context it is required that the tensor network have no open indices, so only positive integer labels are required. The variable \ttt{legLinks} is then a $1\times n$ cell array with each entry being a row vector whose entries are the integer labels associated with the corresponding tensor. The ordering of these labels matches the ordering of the indices on the corresponding tensor in \MATLAB{}. For example, \fref{fig:labelledMERA}(i) shows a closed diagram from the 3:1 1D MERA where all indices have been labelled with positive integers. 
A convention is adopted for relating the diagrammatic indices to indices in \MATLAB{}, whereby the indices of a specific \MATLAB{} tensor are associated with specific legs on the diagram. This is illustrated in \fref{fig:labelledMERA}(ii), where (for example) the topmost leg on tensor~A is associated with the fourth index of the \MATLAB{} tensor \ttt{A(:,:,:,:)}. If tensor~A is the first object to appear in \ttt{tensorList} and tensor~B is the second then, reading the labels associated with tensors~A and~B off the diagram of \fref{fig:labelledMERA}(i), \ttt{legLinks} takes the form \ttt{\{[1 2 3 13],[8 11 12 14],\ldots\}}.

\ttt{sequence} is a row vector comprising positive integers and (optionally) zero, and specifies a contraction sequence for the closed tensor network. Assuming all indices in \fref{fig:labelledMERA}(i) are of identical dimension, denoted $\chi$, an optimal index contraction sequence for this example network is 
\begin{equation*}
\ttt{[11 12 14 15 7 6 5 4 9 8 10 16 1 2 3 13]}, 
\end{equation*}
having a cost of $2\chi^8+2\chi^7+2\chi^6+\chi^4$. Interpretation of this sequence proceeds as follows: The first entry in the sequence is index~11, connecting tensors B and H. The first step of the contraction sequence is therefore to contract together these two tensors, denoted (B,H). This contraction is performed simultaneously over all indices common to both B and H, and therefore also accounts for index~12.\footnote{Note that when two tensors are contracted together, all indices connecting these tensors must be listed consecutively. Sequences which do not satisfy this condition (i)~are always of suboptimal efficiency, and (ii)~are unsupported, and will result in a warning%
.} The next entry in the sequence is index~14, connecting the resulting object to tensor E, indicating that the next contraction is 
\begin{equation*}
\mrm{((B,H),E)}. 
\end{equation*}
Proceeding in this fashion for the entirety of the index list, one obtains the pairwise contraction sequence 
\begin{equation*}
\mrm{((((B,H),E),(((F,G),D),C)),A)}. 
\end{equation*}
Note that if the \ttt{netcon()} reference implementation given in \rcite{pfeifer2013a} is installed then this may be invoked to automatically 
generate an optimal index-based contraction sequence for a tensor network. %

The \ttt{multienv()} function includes support for contraction sequences involving outer products, either where two tensors are contracted despite not sharing an index or where the tensors share only indices of dimension~1, and the appropriate syntaxes for \ttt{sequence} in these situations are discussed in \aref{sec:outerprod}. Further discussion of the index-labelling notation for contraction sequences may be found in \rcite{pfeifer2013a},
including an explicit
algorithm for converting sequences of index labels into sequences of pairwise tensor contractions.

\subsection{Example\label{sec:multienvexample}}

The tensor network given in \fref{fig:labelledMERA} is encountered when variationally optimizing the 3:1 1D MERA in order to compute the ground state of a local Hamiltonian. 
Assuming translation invariance of the Hamiltonian these tensors are taken to satisfy $\mrm{A}=\mrm{B}$, $\mrm{F}=\mrm{C}^\dagger$, and $\mrm{G}=\mrm{H}=\mrm{A}^\dagger$, where $^\dagger$ represents a combination of complex conjugation with vertical reflection in the diagrammatic notation. The tensor which appears in both positions A and B is known as an \emph{isometry} tensor. The environments to be computed from this diagram are:
\begin{enumerate}
\item The environment of E, corresponding to a contribution to the \emph{lifting} of the Hamiltonian (to be returned in \ttt{env1}).
\item The environment of D, corresponding to a contribution to the \emph{lowering} of the reduced density matrix (to be returned in \ttt{env2}).
\item The environments of positions A and B, used in the variational update of the isometry tensor (to be summed and returned in \ttt{env3}).
\item The environment of C, used in the variational update of tensor C (to be returned in \ttt{env4}).
\end{enumerate}
To efficiently compute these environments one may invoke \ttt{multienv()} as follows:
\begin{widetext}
\begin{align*}
&\ttt{tensorList = \{A,A,C,D,conj(C),conj(A),conj(A),E\};}\\
&\ttt{envList = [3 3 4 2 0 0 0 1];}\\
&\ttt{legLinks = \ldots} \\
&\ttt{~~\{[1 2 3 13],[8 11 12 14],[4 9 3 8],[6 5 2 4],[5 9 7 10],[1 6 7 16],[10 11 12 15],[15 16 14 13]\};}\\
&\ttt{sequence = [11 12 14 15 7 6 5 4 9 8 10 16 1 2 3 13];}\\
&\ttt{[env1 env2 env3 env4] = multienv(tensorList,envList,legLinks,sequence);}
\end{align*}
\end{widetext}
The total cost for this invocation is $5\chi^8+4\chi^7+5\chi^6$, which may be compared with a cost of $10\chi^8+10\chi^7+10\chi^6$ if intermediate tensors are not re-used.

\section{Sequences involving outer products\label{sec:outerprod}}

When calculating the environments of tensors in a closed tensor network $\mc{N}$, the need to compute outer products may arise in two distinct situations. The first is where the user-specified contraction sequence for a closed network $\mc{N}$ involves the performance of outer products between two or more tensors appearing in $\mc{N}$. The second is where removal of a tensor~T from a %
network $\mc{N}$ %
causes that network to separate into two or more disjoint parts (or, if $\mc{N}$ is already disjoint, causes a non-disjoint subnetwork of $\mc{N}$ to separate into two or more disjoint parts). Calculation of the environment of~T then necessarily involves taking the outer product of these parts. 
These two situations are not mutually exclusive, and calculation of a tensor environment may include both user-specified outer products and outer products which arise when removing individual tensors to compute their environments.

The second situation---where outer products arise upon deleting a tensor~T from a closed network $\mc{N}$---poses no challenges beyond those already discussed within this paper, and is handled automatically by \ttt{multienv()}. This Appendix, on the other hand, concerns itself with the syntaxes whereby a user may specify 
a contraction sequence to \ttt{multienv()} %
which explicitly involves the performance of outer products.

\subsection{Disjoint networks\label{sec:disjOP}}

When invoking \ttt{multienv()} the supplied tensor network $\mc{N}$ may be disjoint, being made up of $n$ separate closed subnetworks $\mc{N}_1,\ldots,\mc{N}_n$. The problem of contracting network $\mc{N}$ to a number may then be subdivided into the problems of contracting each subnetwork to a number, and multiplying these numbers together. As a number may be understood as a tensor of dimension~0, formally these final multiplications correspond to outer products between tensors of dimension~0. If \ttt{multienv()} is supplied with a disjoint network $\mc{N}$ and a contraction sequence which reduces all subnetworks to numbers, the contraction tree for network $\mc{N}$ is then completed by performing the pairwise contraction (multiplication) of these numbers until only a single numerical value remains. Contraction sequences for the calculation of individual tensor environments are then determined in the normal way by manipulating the resulting fusion tree for the whole of $\mc{N}$ as described in \sref{sec:contcosts}.

\subsection{Trivial indices\label{sec:dim1}}

\begin{figure}
\includegraphics[width=246.0pt]{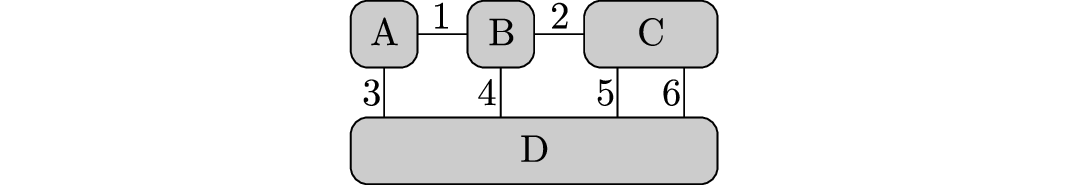}
\caption{Example tensor network with trivial index. All indices have dimension~3 except for the index labelled 2, which has dimension~1.\label{fig:trivindexample}}
\end{figure}

To perform outer products between tensors of dimension greater than zero during the contraction of a tensor network $\mc{N}$, one may introduce labelled connecting indices of dimension~1. Consider the example %
\begin{align*}
&\ttt{A = rand(3,3); B = rand(3,1,3);}\\
&\ttt{C = rand(3,3,1); D = rand(3,3,3,3);}\\
&\ttt{tensors = \{A,B,C,D\};}\\
&\ttt{envList = [0 0 0 1];}\\
&\ttt{legs = \{[3 1],[1 2 4],[5 6 2],[3 4 5 6]\};}\\
&\ttt{seq = [1 2 3 4 5 6];}\\
&\ttt{envD = multienv(tensors,envlist,legs,seq);}
\end{align*}
where the dimensions of tensors~B and~C are $3\times 1\times 3$ and $3\times3\times1$ respectively.\footnote{Note that \MATLAB{} leaves trailing indices of dimension~1 implicit, and so will report the size of tensor~C as $3\times 3$ if queried, and not $3\times 3\times 1$. Equally, it is also valid to create tensor~C using the command \ttt{C~=~rand(3,3);} instead of \ttt{C~=~rand(3,3,1);} as given above.} %
This tensor network is illustrated in \fref{fig:trivindexample}, 
and the given index contraction sequence corresponds to a pairwise tensor contraction sequence of (((A,B),C),D). 
The outer product %
is between %
the contraction product (A,B) and the tensor~C, and corresponds to contraction over the index of dimension~1 carrying the numerical %
label \ttt{2}. This label appears in the index-based contraction sequence in the usual manner.

The inclusion of trivial indices %
is the most versatile means of specifying an outer product %
as part of the contraction sequence%
, with the only drawback being the need to explicitly include indices of dimension~1 over which the contractions are to be performed.

\subsection{Zeros-in-\ttt{sequence} notation\label{sec:zis}}

While any contraction sequence involving outer products may always be described using the trivial index approach of \aref{sec:dim1}, it is sometimes useful to use a different notation involving the insertion of zeros into the index sequence passed to \ttt{multienv()}. This notation was introduced in \rcite{pfeifer2013a} as a means of describing contraction sequences involving outer products when appropriate trivial indices are not present. Although this notation is not capable of describing every contraction sequence which involves an outer product, for any given tensor network there always exists an optimal (minimal cost) contraction sequence which can be described using this notation.

\begin{figure}
\includegraphics[width=246.0pt]{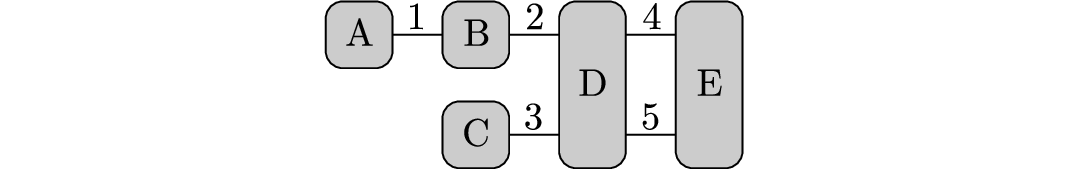}
\caption{Example tensor network; all indices have dimension~2. An index sequence of \ttt{[1 0 2 3 4 5]} corresponds to the pairwise tensor contraction sequence (((A,B),C),D),E) where the second contraction to be performed is an outer product.\label{fig:zeroex}}
\end{figure}

In brief, in \rcite{pfeifer2013a} it was shown that for any tensor network, if an outer product of two or more tensors is \emph{required} as part of the optimal contraction sequence and the result of this outer product is denoted X, then an optimal contraction sequence may always be found where this outer product is always either
\begin{enumerate}
\item the final step in the contraction of the tensor network, or
\item followed by contracting \emph{all} indices of object~X with another tensor.
\end{enumerate}
Outer products of these forms (and these forms only) may be represented %
by inserting zeros into the contraction sequence, with the outer product of $n$ tensors being indicated by $n-1$ consecutive zeros. To determine the $n$ tensors involved in the outer product when there are more than $n$ tensors remaining, indices are read from the sequence after the zeros, and the tensors carrying these indices are noted, until $n+1$ tensors have been identified. Given the above constraints on the outer products which may be represented using this notation, it then follows hat %
one of these $n+1$ tensors will necessarily share summed indices with all $n$ other tensors. This tensor does not participate in the outer product, but is instead the tensor which is subsequently contracted with object~X. For a fuller discussion of this outer product notation, and of the optimal performance of the outer products of multiple tensors, see \rcite{pfeifer2013a}. %
An example is given by
\begin{align*}
&\ttt{A = rand(2,1); B = rand(2,2); C = rand(2,1);}\\
&\ttt{D = rand(2,2,2,2); E = rand(2,2);}\\
&\ttt{tensors = \{A,B,C,D,E\};}\\
&\ttt{envList = [0 0 0 0 1];}\\
&\ttt{legs = \{[1],[1 2],[3],[2 3 4 5],[4 5]\};}\\
&\ttt{seq = [1 0 2 3 4 5];}\\
&\ttt{envE = multienv(tensors,envList,legs,seq);}
\end{align*}
where the index sequence \ttt{[1 0 2 3 4 5]} corresponds to a tensor contraction sequence ((((A,B),C),D),E) and the network is illustrated in \fref{fig:zeroex}. %

While this approach is not as versatile as the use of trivial indices discussed in \aref{sec:dim1}, it is nevertheless useful when an optimal contraction sequence is being automatically generated, for instance by using the Netcon algorithm.\cite{pfeifer2013a} Because there always exists an optimal contraction sequence which may described using the zeros-in-\ttt{sequence} notation, a search algorithm for optimal sequences can return a valid response in this notation even when all optimal contraction sequences %
involve %
performing 
an outer product and no appropriate 
index of dimension~1 has been provided. 

By supporting the zeros-in-\ttt{sequence} notation for outer products, \ttt{multienv()} is capable of directly accepting contraction sequences generated by the reference implementation of Netcon provided with \rcite{pfeifer2013a}.

\bibliography{TensorProof}

\end{document}